\documentclass{aims_LENCOS}
\usepackage{amsmath}
  \usepackage{paralist}
  \usepackage{graphics} 
  \usepackage{epsfig} 
 \usepackage[colorlinks=true]{hyperref}
\hypersetup{urlcolor=blue, citecolor=red}

\def\currentvolume{X}
 \def\currentissue{X}
  \def\currentyear{200X}
   \def\currentmonth{XX}
    \def\ppages{X--XX}

\theoremstyle{definition}

\title[Analysis of Supercontinuum Generation]
      {Analysis of Supercontinuum Generation Under General Dispersion
Characteristics and Beyond the Slowly Varying Envelope Approximation}

\author[A. B. Aceves, Y. Chung, T. Hagstrom, R. Chen and M. Hummel]{}


 \email{aaceves@smu.edu}

\thanks{A. B. Aceves, Y. Chung and R. Chen were supported by NSF Grant DMS-0505618,
T. Hagstrom was supported by NSF Grant DMS-0610067}

\begin{document}
\maketitle

\centerline{\scshape Alejandro B. Aceves}
\medskip
{\footnotesize
 \centerline{$^1$ Department of Mathematics}
   \centerline{Southern Methodist University}
   \centerline{Dallas TX 75275, USA}}

\medskip

\centerline{\scshape Rondald Chen$^2$, Yeojin Chung$^1$, Thomas Hagstrom$^1$ and Michelle Hummel$^2$}
\medskip
{\footnotesize
 \centerline{$^2$Department of Mathematics and Statistics}
   \centerline{The University of New Mexico}
   \centerline{Albuquerque, NM 87131, USA}
}

\bigskip


\begin{abstract}
The generation of broadband supercontinua (SC) in air-silica
microstructured fibers results from a delicate balance of dispersion and
nonlinearity. We analyze two models aimed at better understanding SC.
In the first one, we
characterize linear dispersion in the Fourier domain from the calculated
group velocity dispersion (GVD) without using a Taylor approximation for the
propagation constant. Results of our numerical
simulations are in good agreement with experiments. A novel relevant length
scale, namely the length for shock formation, is introduced and its role
 is discussed. The second part shows
similar dynamics for a model that goes beyond the slowly varying
approximation for optical pulse propagation.
\end{abstract}


\section{Introduction}\label{S:intro}
Since the experimental demonstration of optical soliton propagation
in single mode fibers some 20 plus years ago, the investigation of
pulse dynamics in nonlinear
optical fibers has evolved due to the introduction of novel structures
with complex properties,
such as photonic crystal and holey fibers \cite{kba96}.
In essence these are examples
of engineered dielectric structures aimed at tailoring dispersive
characteristics and enhancing nonlinear behavior. A direct outcome in terms
of the pulse dynamics that has brought much attention from several
experimental groups \cite{dudley1,genty,dudley,ds02,bwr00,wok02,sol06,
kgr02,hgz02,oww06}  is the ability to generate broadband supercontinuum
spectra. Scientifically
this is a departure from soliton dynamics that requires
careful analytical and numerical modeling in parallel with the
experiments. From the applications point of view, it has opened
possibilities never seen before in areas such as frequency metrology
\cite{uhh02} and medical diagnostics \cite{hlc01,syl04,lks04}.

Supercontinuum generation (SCG) can arise from various physical processes such
as self- and cross-phase modulation, and amplitude modulation \cite{alfano}.
Due to the complex interplay of
linear and nonlinear phenomena in SCG dynamics, the
theoretical formulation of the
SCG mechanism imposes considerable challenges, in particular if this
process happens in bulk media \cite{moloney}.
The major recent theory that explains the SCG for
relatively low intensities in confined waveguides rests on the evolution and
fission of higher-order solitons near the
zero-dispersion wavelength in PCFs \cite{hgz02,hh01,dgxkzstcw02}. If the input wavelength is
close to the zero-dispersion wavelength, then the influence of third-order dispersion is strong,
thus a higher-order soliton with number $N$ splits into its constituent solitons with the
emission of blueshifted nonsolitonic radiation \cite{wmlc86}. Since each soliton and its
corresponding radiation  has a different central frequency, the width of the generated total
spectrum increases with increasing soliton number.

Recent experimental observations of supercontina in soft glass (schott SF6)
PCF \cite{kgr02,oww06}, however,
suggest an interesting physical mechanism of SCG that cannot be fully explained by the previously
known theories. In these experiments, SCG occurs in a dramatic fashion in the very early
states of propagation, in particular at a length scale where solitons start forming.
Such a phenomenon can only be
explained if, initially, nonlinear effects other than soliton fission dominate the physics.
Indeed, the underpinning mechanisms that generate supercontinua as reported in most theoretical and
experimental studies are shock generation and its dispersive regularization in combination with
multisoliton fission. The shock generation is a well known classical phenomenon in fluids and
gas dynamics \cite{whitham}. It also
appears in ultra-short pulse propagation in fibers \cite{agrawal,anderson,ys84}. For ultrashort pulses, the
refractive index depends on the pulse intensity, thus the center of the pulse envelope travels
with a
different speed than that of the trailing and leading edges of the pulse;
this leads to an
asymmetric
shape of the pulse, which invokes shock formation. However, in optical propagation,
dispersion plays an important role, preventing a sharp discontinuity.
On the other hand, multisoliton generation resulting
from small dispersion effects is a consequence of the integrability of
the nonlinear Schr\"odinger equation (NLSE)  \cite{zakharov}.
Its eventual fission is
the result of perturbations to the NLSE such as third order
dispersion. Altogether, a universal feature of nonlinear dispersive wave
phenomena is that
the long term dynamics results from the delicate balance between
linear and nonlinear effects.

To better illustrate this delicate balance we begin by studying a simpler model. Here
we do not aim at studying a particular fiber; instead we highlight with a minimal number of linear
and nonlinear terms different output scenarios. For this, consider the equation
\begin{equation}\label{NLSE}
i\partial_z A + c_3 \partial^2_{t} A + i \epsilon \partial^3_{t} A
+ |A|^2 A + ic_2 \partial_t (|A|^2A) =0,
\end{equation}
where $A$ is the envelope of the pulse, $c_3, \epsilon$ are
respectively coefficients for the second and third linear dispersion
terms, and $c_2$ is the coefficient of a self-steepening term. We can then
characterize the dynamics of pulses under different regimes.
In all instances we assume an input pulse $A_{inp} = \sqrt{2}/\cosh (t)$.

\noindent
{\bf (i)} $0 < \epsilon,  c_2, c_3 \ll 1$ for which
$N \approx \frac{\sqrt{2}}{\sqrt{c_3}}$ solitons are created and
third order dispersion and self-steepening are viewed as perturbations
that split the solitons (Figure 1).\\

\begin{figure}[htp]
\begin{center}
\includegraphics[width=4in]{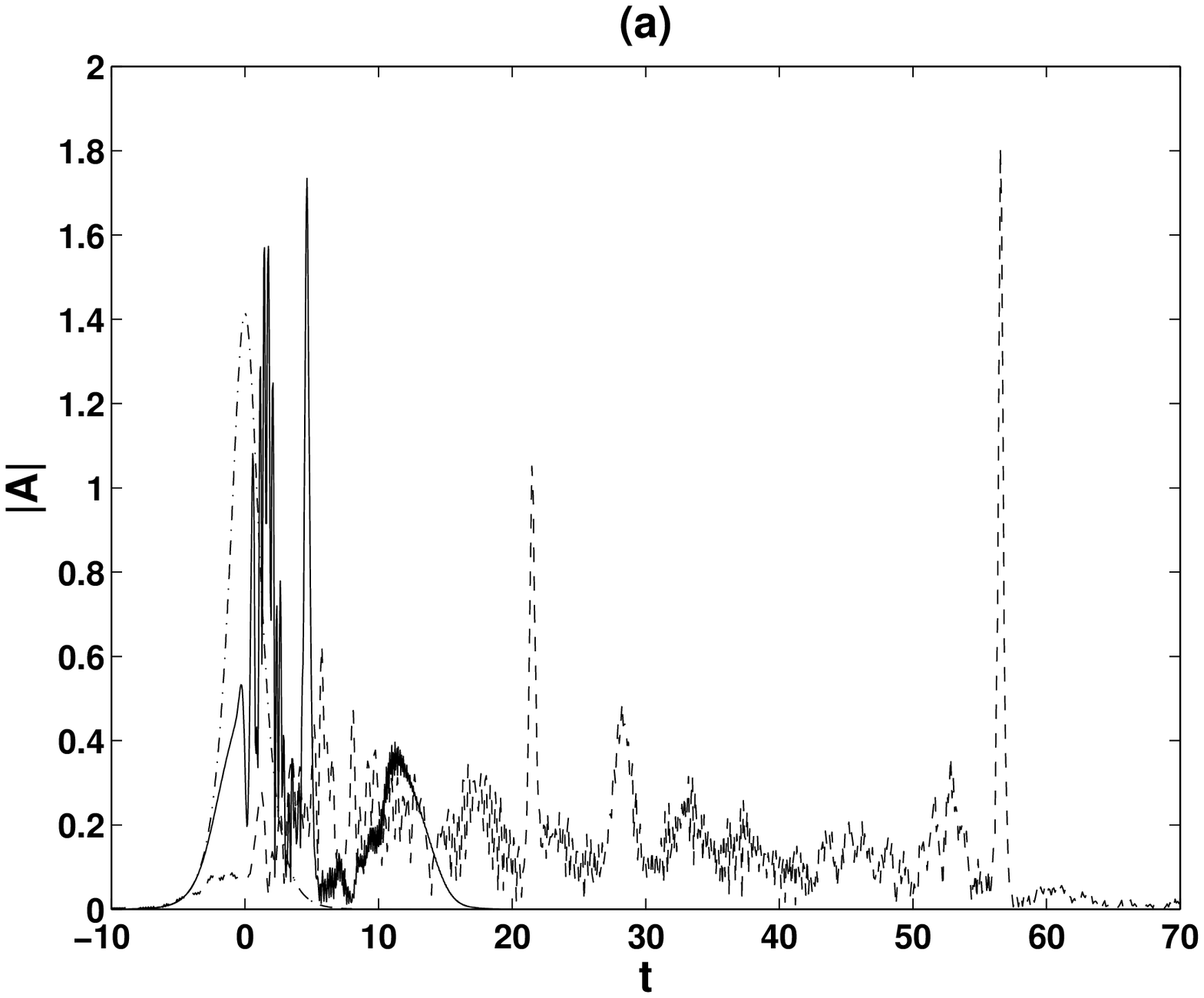}
\includegraphics[width=4in]{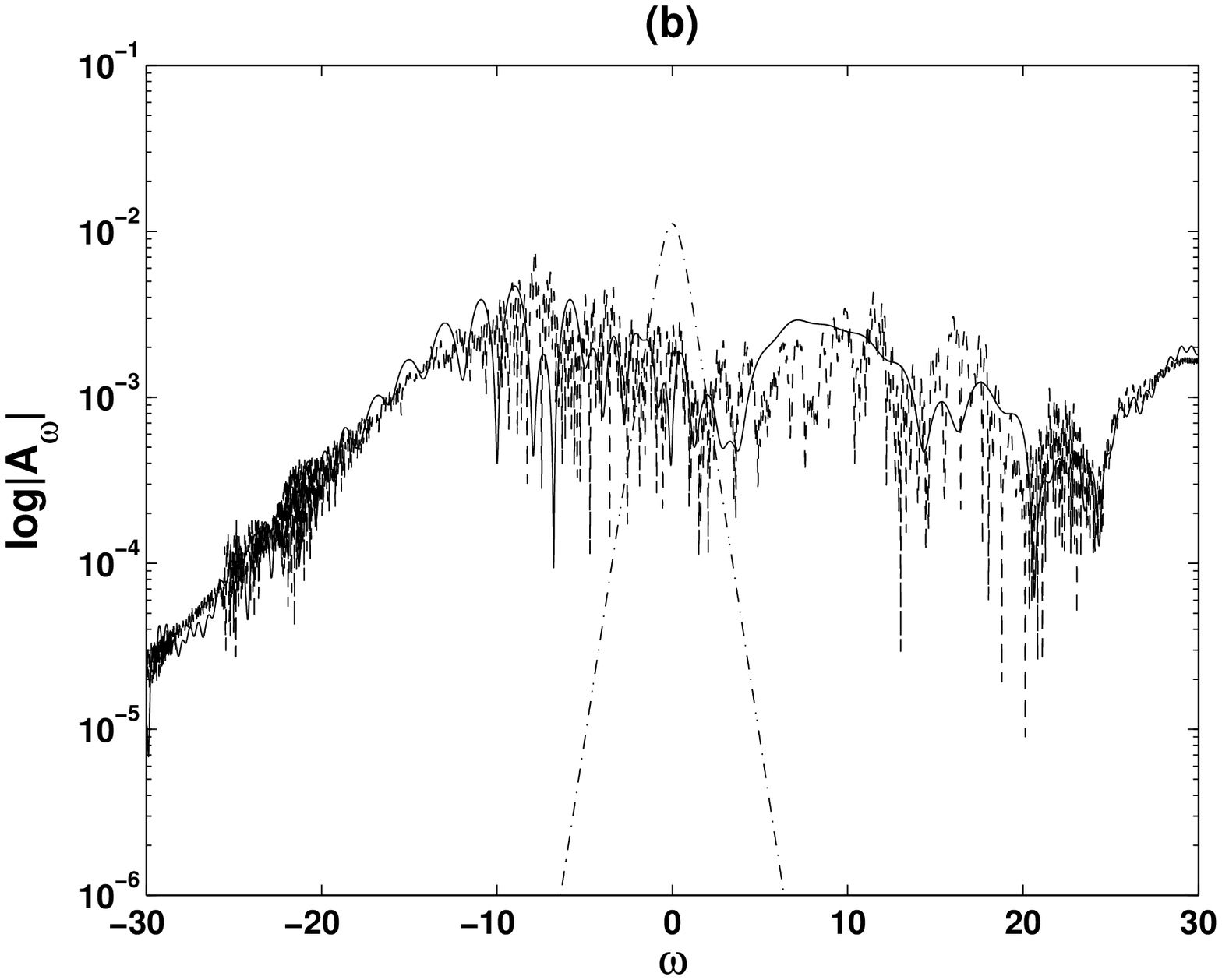}\\
\caption{Spectral (bottom)
and temporal (top) picture of the pulse profile at $z=0$ (dash-dotted),
$z=10$ (solid) and
at $z=100$ (dashed).
Parameter values $\epsilon = 0.01, c_2 = 0.001, c_3 = 0.01$.}
\label{F:figure1}
\end{center}
\end{figure}

\noindent
{\bf (ii)} $0< \epsilon,  c_3 \ll 1$ and both much smaller than $c_2$,
for which the dominant effect is shock formation at a propagation length
proportional to $\frac{1}{2c_2}$. The dispersion terms are viewed as
perturbations which may or may not prevent shock formation. If they do, one
expects spectral broadening (Figure 2). \\

\begin{figure}[htp]
\begin{center}
\includegraphics[width=4in]{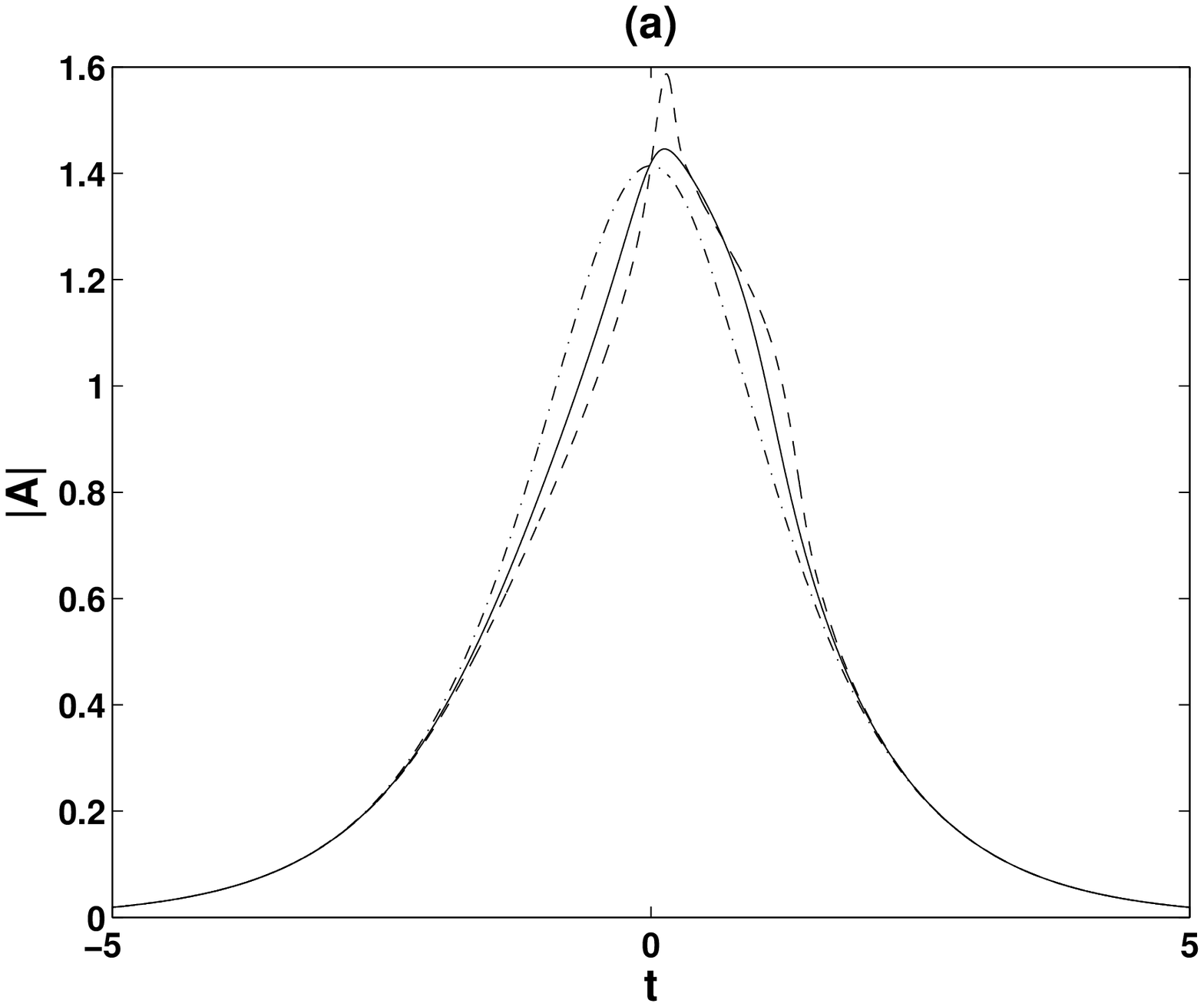}
\includegraphics[width=4in]{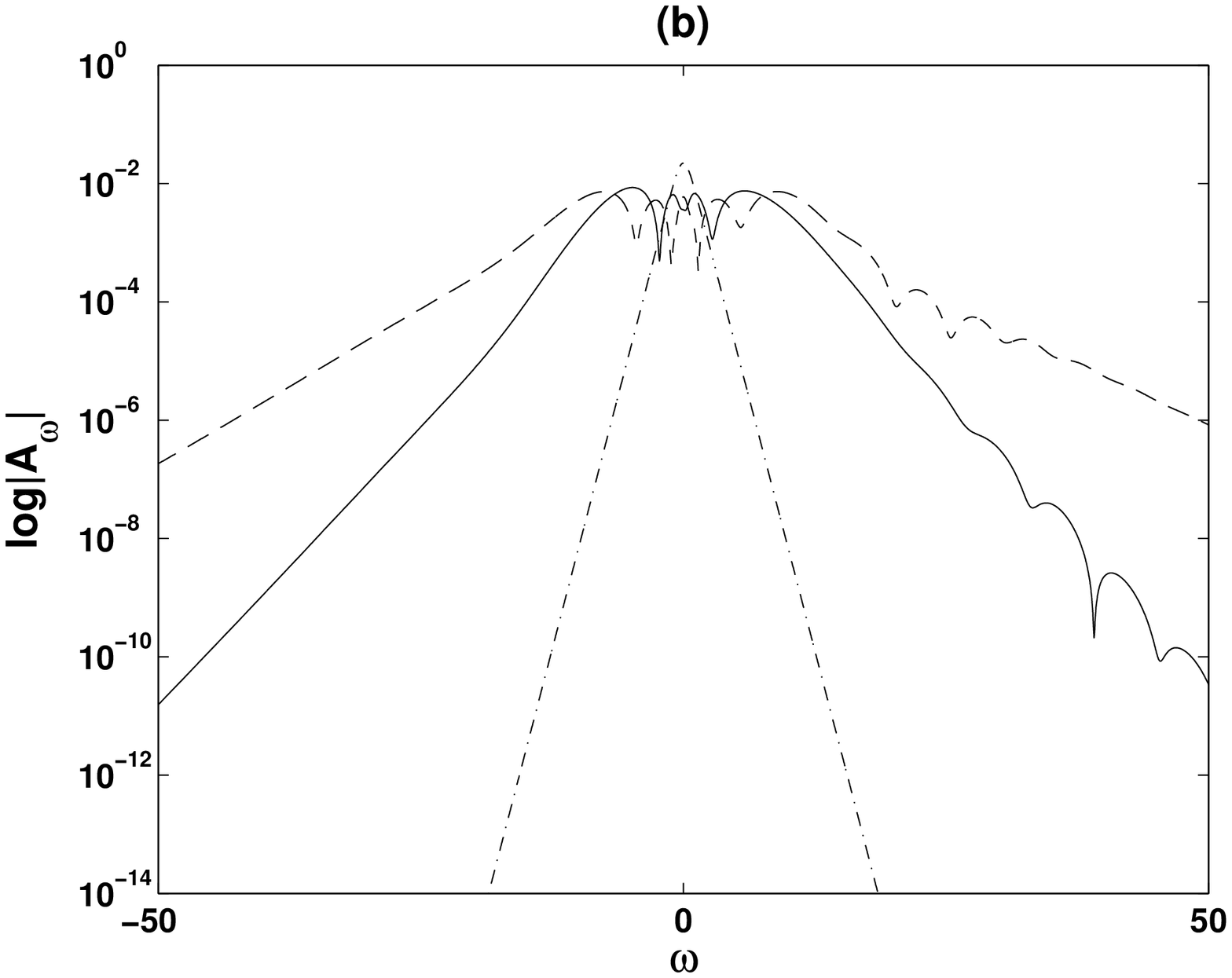}\\
\caption{Spectral (bottom)
and temporal (top) picture of the pulse profile at $z=0$ (dash-dotted),
$z=5$ (solid) and
at $z=7$ (dashed). Parameter values $\epsilon = 0.0001,
c_2 = 0.01, c_3 = 0.0004$.}
\label{F:figure2}
\end{center}
\end{figure}

In contrast, if linear dispersion is dominant and of order one, two cases
which illustrate the overall dynamics are, first:

\noindent
{\bf (iii)} $c_3 = 1, \ c_2 \ll \epsilon \ll 1$ for which a soliton emerges and its dynamics under
higher order corrections is well described by soliton perturbation theory
\cite{biswas} (Figure 3). In addition,
third order dispersion accounts for linear wave resonances which explain the peak in the spectrum
(figure 3b).

\begin{figure}[htp]
\begin{center}
\includegraphics[width=4in]{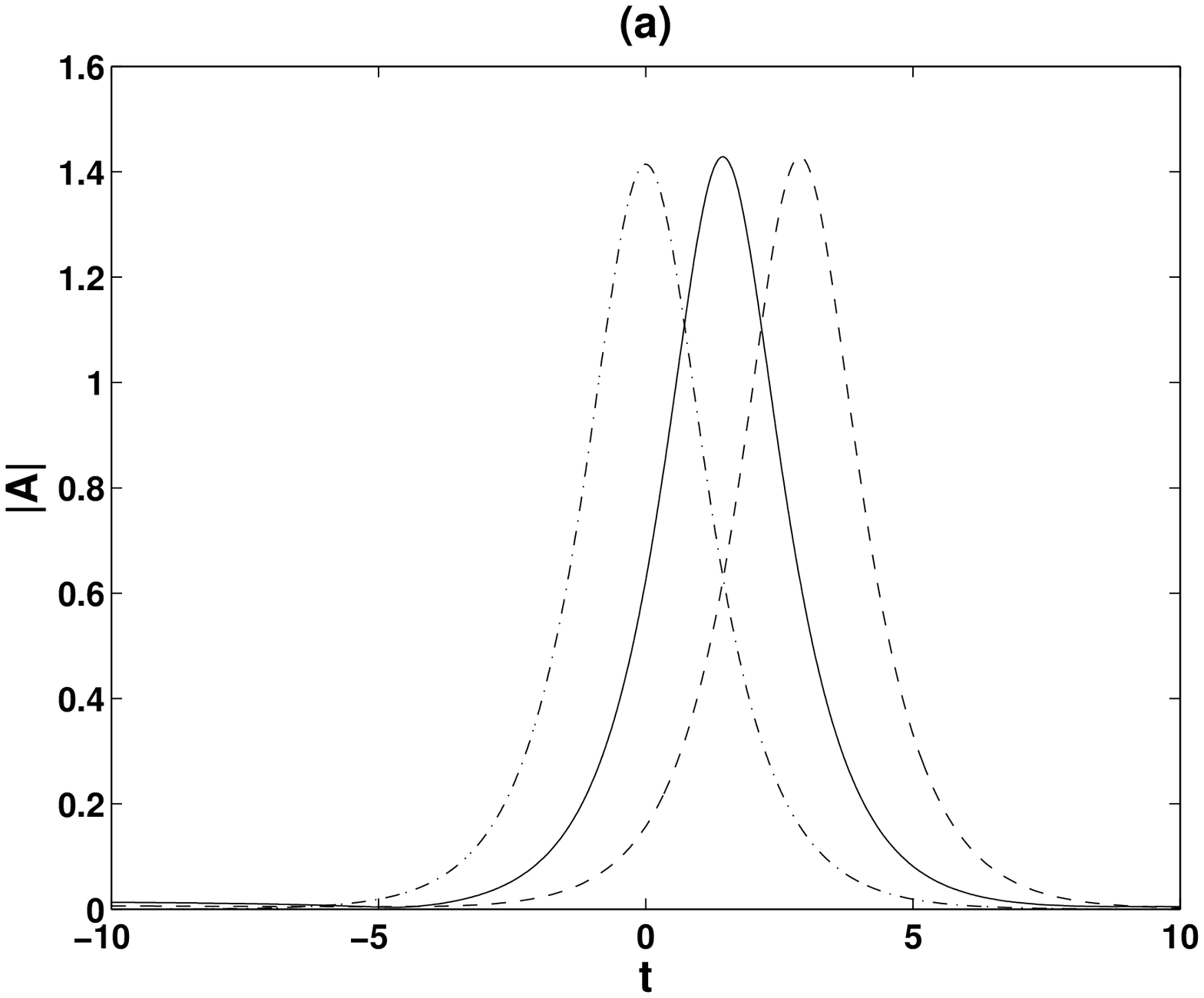}
\includegraphics[width=4in]{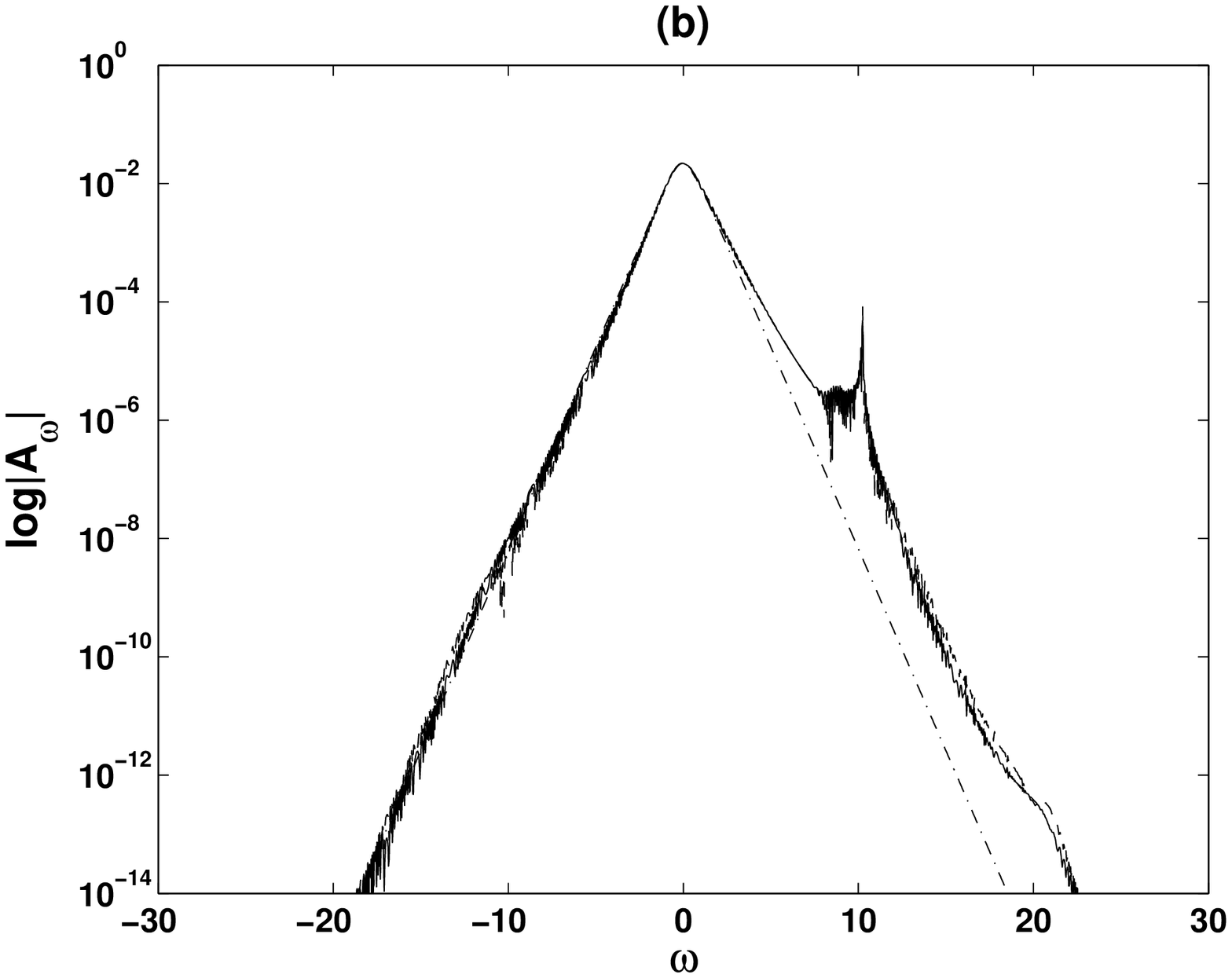}\\
\caption{Spectral (bottom)
and temporal (top) picture of the pulse profile at $z=0$ (dashed-dotted),
$z=10$ (solid) and at $z=20$ (dashed). Parameter values
$\epsilon = 0.1, c_2 = 0.02, c_3 = 1.0$.}
\label{F:figure3}
\end{center}
\end{figure}

\noindent
Finally,

\noindent
{\bf (iv)} $\epsilon = 0.1, \ c_3 \ll c_2 \ll 1$ for which there is no soliton (Figure 4).

\begin{figure}[htp]
\begin{center}
\includegraphics[width=4in]{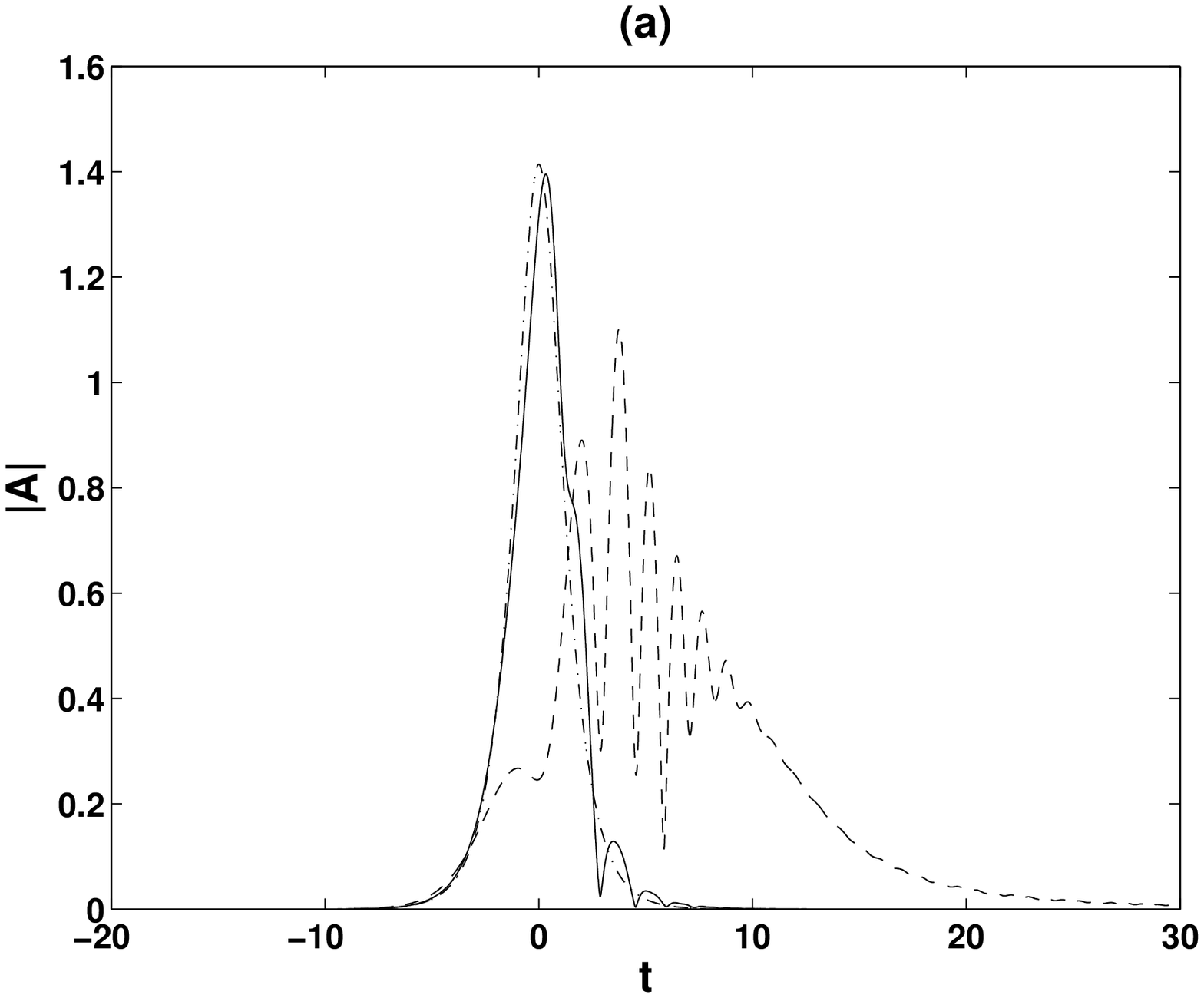}
\includegraphics[width=4in]{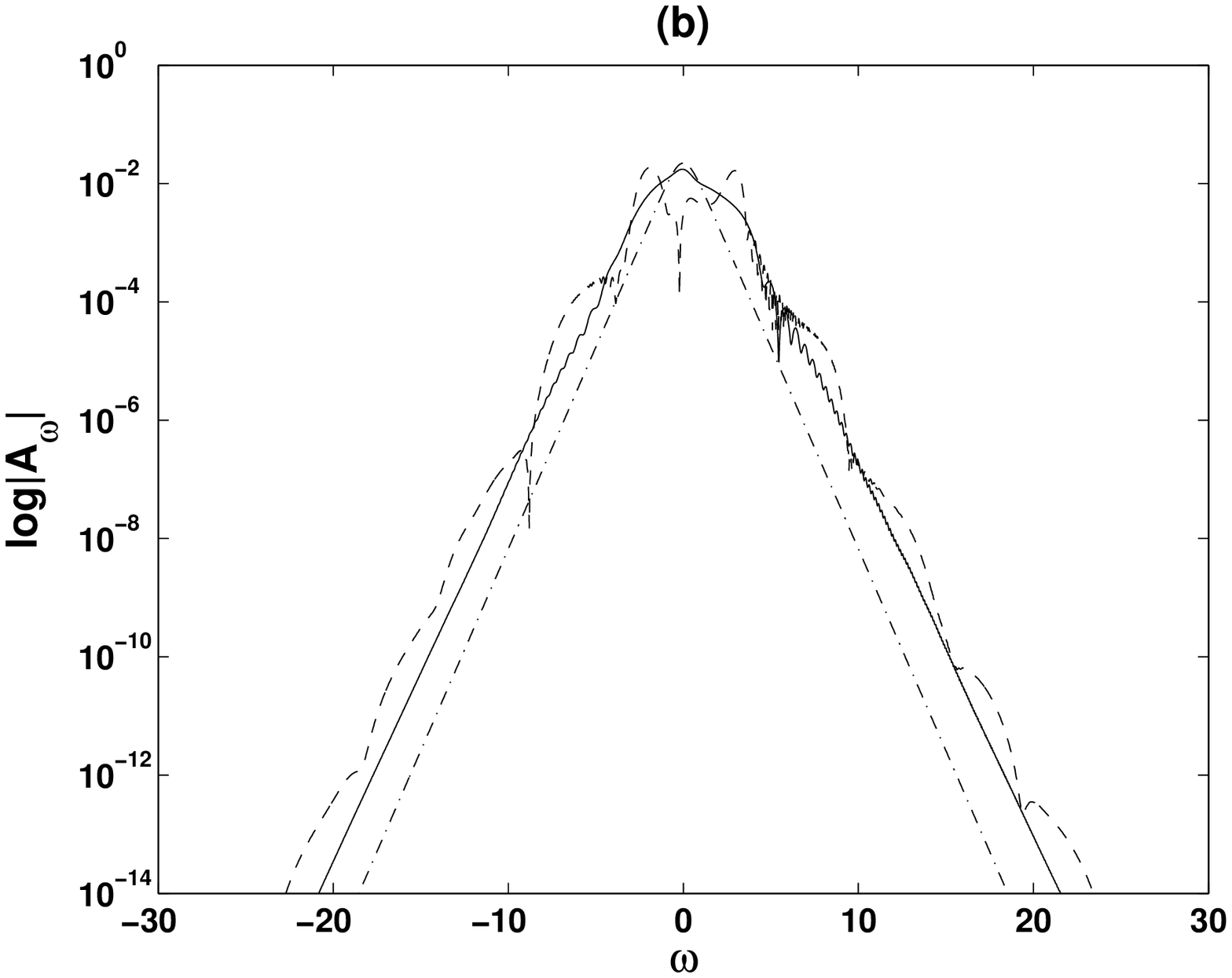}\\
\caption{Spectral (bottom)
and temporal (top) picture of the pulse profile at $z=0$ (dashed-dotted),
$z=1$ (solid), $z=5$
(dashed). Parameter values $\epsilon = 0.1, c_2 = 0.01, c_3 = 0.0001$.}
\label{F:figure4}
\end{center}
\end{figure}

Notice the striking difference in the spectrum at the output. While a broadening of the
spectrum is achieved in the first two scenarios, this is not the case for the last two, even if,
as the temporal profile displayed in case (iv) suggests, the initial pulse is destroyed.

In the next section, we recognize the aformentioned outcomes for a more realistic model
describing the pulse dynamics in photonic
crystal fibers. The extended model
accounts for all competing effects including self-steepening, which we believe is as
important as the effect from the fully detailed linear dispersion. As we hoped the first
part showed, in order to understand  and exploit these phenomena,
it is essential to obtain and analyze better these mathematical models. This in addition could
explain for each instance in a real experiment what triggered SC generation. To begin with,
the accurate broadband modeling of the
dispersion relation is required to make sure one does not obtain
spurious results, and to do so here we depart from the
commonly used approach where a Taylor series expansion of the propagation constant $\beta$ models
the dispersive properties in a generalized nonlinear Schr\"odinger equation
(gNLSE). Instead, we develop a mathematical model starting from
calculated group velocity dispersion (GVD) curves. Then, we construct the function
$\beta(\omega)$ over a broad frequency window
and integrate the gNLSE preserving the spectral dependence of the
propagation constant. As an illustration, we present our numerical results
based on the calculated GVD for an $LP_{01}$ mode in an air-silica
microstructured fiber studied by Dudley \emph {et al}. \cite{dudley}.
Then, similar to what we did with equation (1), we carry out a careful numerical analysis.
We find that if the nonlinear self-steepening term is strong enough, the model as it
stands produces a shock that is not arrested by dispersion, whereas for weaker nonlinearity
the pulse propagates the full extent of the fiber with the generation of a supercontinuum.

Recent studies, in particular for ultrashort pulse dynamics \cite{CS07}, have recognized that the
slowly varying envelope approximation may not hold as a good model. In the last part of this paper we
discuss how spectral broadening arises without invoking the slowly varying envelope approximation (SVEA).

\section{Formulation of the model}\label{model}
The propagation of an electric field wave packet through an optical
fiber can be described by gNLSE \cite{agrawal},
\begin{equation}\label{gNLSE_general}
i\partial_z A +{\mathcal F}^{-1}[(\beta(\omega)-\beta(\omega_0))\hat{A}] + \gamma\left(1 +
\frac{i}{\omega_0}\partial_t\right)(A|A|^2)   =0.
\end{equation}
\noindent
Here, the variables $z,T,\omega$ represent propagation distance, time and optical frequency, respectively.
The envelope of the wave packet is $A$, and $c, \lambda,\omega_0, \beta(\omega)$ represent
the velocity of light in vacuum, wavelength, central frequency and wave number, respectively.
$\mathcal{F}^{-1}$ denotes the inverse Fourier transform, and $\hat{A}$ is the Fourier transform of the pulse
envelope.
Finally the self-steepening term models the instantaneous nonlinear response function of the medium, which
is a good approximation given the temporal lengths of the pulses. The inclusion of a Raman (non-instantaneous)
term we believe will only introduce a shift in  the peak location in the spectrum (see Fig.~\ref{F:spectrum1}
and compare with Fig.~5b in \cite{dudley}).

The effects of fiber dispersion are accounted for by the propagation constant $\beta(\omega)$ which we
calculate based on the dispersion profile presented in \cite{dudley}, without performing a
Taylor series expansion
around the carrier frequency. Using two high-precision numerical
integrations of an accurate rational
interpolant of the GVD curve, we obtain the GVD function $D(s)$. Then, the group
velocity $\nu_g(\omega)$
is derived from $D(s)$ through the relation,
\begin{equation}
\frac{1}{\nu_g(\omega)} - \frac{1}{\nu_g(\omega_0)} = \int_{\lambda_0}^{\lambda} D(s) ds.
\end{equation}
By setting $F(\lambda) = \int_{\lambda_0}^{\lambda} D(s) ds$, we obtain
\begin{equation}
\nu_g(\omega) = \frac{\nu_{g}(\omega_0)}{1+\nu_g(\omega_0)F(\lambda)}.
\end{equation}
Since $\frac{\partial \beta}{\partial \omega} = \frac{1}{\nu_g} $, it follows that
\begin{equation}\label{eq1}
\frac{\partial \beta}{\partial \omega} = \frac{1}{\nu_g(\omega_0)} + F(\lambda) \sim \frac{1}{c} + F(\lambda).
\end{equation}
By integrating Eq.~(\ref{eq1}) with respect to $\omega$ and using the relation $\lambda = \frac{2\pi c}{\omega}$,
we obtain
\begin{equation}
\beta(\omega)-\beta(\omega_0) = \frac{\omega-\omega_0}{\nu_g(\omega_0)} -
2\pi c\int_{\lambda_0}^{\lambda}\frac{F(\lambda)}{\lambda^2}d\lambda.
\end{equation}
We employ a frame of reference moving with the pulse at the group velocity $\nu_g$ by making the
transformation $t = T-z/{\nu_g}$. In the end, we obtain
\begin{equation}\label{gNLSE}
i\partial_z A - 2\pi c{\mathcal F}^{-1}\left(
\int_{\lambda_0}^{\frac{2\pi c}{\omega}} \frac{F(\lambda)}{\lambda^2}
d\lambda \hat{A}(\omega, z)\right) + \gamma\left(1 +
\frac{i}{\omega_0}\partial_t\right)(A|A|^2)   =0.
\end{equation}
\noindent

The resulting equation preserves the complete structure of fiber dispersion which is indeed utilized in
experiments. In addition, the equation is valid not only for broad pulses, but also short pulses since the
derivation is carried out without the assumption of a pulse centered around a specific carrier frequency
(without Taylor series expansion of the propagation constant around a
carrier frequency). In the remainder of this
section, we present analytical and numerical results obtained from the gNLS (7).

\subsection{Optical shock formation} \label{shock}
In order to first pay attention to the nonlinear effects governing the
mechanism of shock formation \cite{agrawal, anderson},
we consider the dispersionless case by setting $F(\lambda)=0$ in Eq.~(\ref{gNLSE}).
In the absence of dispersion, we first split Eq.~(\ref{gNLSE}) into
an intensity-phase system by adding and subtracting
\begin{eqnarray}
{A^*} \frac{\partial A}{\partial z} &=&-\frac{\gamma {A^*}}{\omega_0}\frac{\partial}{\partial t}(|A|^2 A) + i\gamma |A|^4,
\label{shock1}\\
{A} \frac{\partial {A^*}}{\partial z} &=& -\frac{\gamma A}{\omega_0}\frac{\partial}{\partial t}(|A|^2 A^*)- i\gamma |A|^4,
\label{shock2}
\end{eqnarray}
where $A^*$ is the complex conjugate of $A$.

By defining $I= |A|^2$, the addition of Eqs.~(\ref{shock1}), \ (\ref{shock2}) gives
\begin{eqnarray}
\frac{\partial I}{\partial z} &=&  -\frac{\gamma}{\omega_0}\left[2|A|^2
\frac{\partial |A|^2}{\partial t} +
|A|^2 \left( A^* \frac{\partial A}{\partial t} + A \frac{\partial A^*}{\partial t}\right)\right]
\nonumber\\
&=& -\frac{3 \gamma}{\omega_0}I \frac{\partial I}{\partial t}. \label{intensity}
\end{eqnarray}
The general solution of Eq.~(\ref{intensity}) is
\begin{equation}
I(z,t) = f\left(t-\frac{3\gamma}{\omega_0}Iz\right), \label{intensity_sol}
\end{equation}
where $f(t)$ is determined by the initial pulse shape, namely, $f(t) = I(0,t)$.
The solution form Eq.~(\ref{intensity_sol}) implies that asymmetric
distortion of the pulse will occur eventually.

From Eq.~(\ref{intensity_sol}), we also find
\begin{equation}
\frac{\partial I}{\partial t} = \frac{f'}{1+\frac{3\gamma}{\omega_0}f'z}.
\end{equation}
The resulting equation shows that after a distance $z_s = -\frac{\omega_0}{3\gamma}\frac{1}{f'}$,
a singularity in the pulse intensity will be generated, namely, the formation of an optical shock. This
shock does play an important role in the
spectral broadening once dispersion regularizes it.
In other words, the effects of fiber dispersion cannot be ignored.
Moreover, the effect of GVD becomes more important as the pulse steepening
becomes significant. This phenomenon
prevents further steepening of pulse shape, i.e.,an appropriate strength of linear
dispersion results in a mechanism that may prevent (or regularize) the shock.

\subsection{Numerical solutions of the
generalized nonlinear Schr\"odinger equation}

Using the GVD profile employed in experiments \cite{dudley}, we perform our numerical simulations of pulse
dynamics based on Eq.
~(\ref{gNLSE}). In particular, we consider the propagation of 100$fs$ pulses at 780$nm$ in a 1-m length
air-silica microstructured fiber with $\gamma=0.1W^{-1}m^{-1}$. We assume that the input pulse has a form of
$A_0^2/\cosh^2\frac{t}{t_0}$. As a reference, we find from the previous analysis
that the shock length for this input pulse is $z_s = 22.1cm$ which is much shorter than the actual
fiber length, thus in this case shock regularization is a likely scenario for SC generation.

The integration spectrally covers the $400-900 nm$ range and, as stated above, it does not
require a Taylor expansion of $\beta$. Instead we calculate $\beta$ via two
high-precision numerical integrations of the GVD. The numerical evolution is then completed using a
standard adaptive ode solver. The results displayed below (figures 5, 6)
are in clear qualitative agreement with those in \cite{dudley}. We should point out two important distinctions
between figure 5  in \cite{dudley} and figure 5 here: we do not capture the peak in the spectrum at wavelengths
close to $1200nm$ in 5a,b of \cite{dudley} and the corresponding pulse (labeled $C$) shown in figure 5c. This is
because in our approach we computed the dispersion profile based on the calculated GVD curve shown in figure 2 of
\cite{dudley}. This calculation did not extend to wavelengths beyond $900nm$ and we did not extrapolate such curves.
This explains the sharp decay in the spectrum of figure 5 here. Calculations based on a Taylor expansion of
$\beta$ which is commonly used, can in principle be extended to any desired spectral range. On the other hand, our
result better reproduces the observed supercontinuum spectrum (figure 5a of \cite{dudley}) in the short (less than
$600nm$) wavelength portion and is as good as the Taylor expansion in the intermediate regime.

Finally, figure 5 (right)  shows five distinguishable pulses at the
output. In figure 6, we spectrally isolate each pulse and find their peaks
centered approximately at: $867nm$ (peak 1), $913nm$ (peak 2),
$848nm$ (peak 3), $840nm$ (peak 4)   $858nm$ (peak 5). As we state below,
the spectral separation should be accentuated by the presence of the Raman
shift which we did not include in the model. What is most important here is we
corroborate spectral broadening  and that splitting of pulses occurs, with the spectral
shift accounting for differences in soliton velocities.

\begin{figure}[htp]
\begin{center}
\includegraphics[width=4.5in]{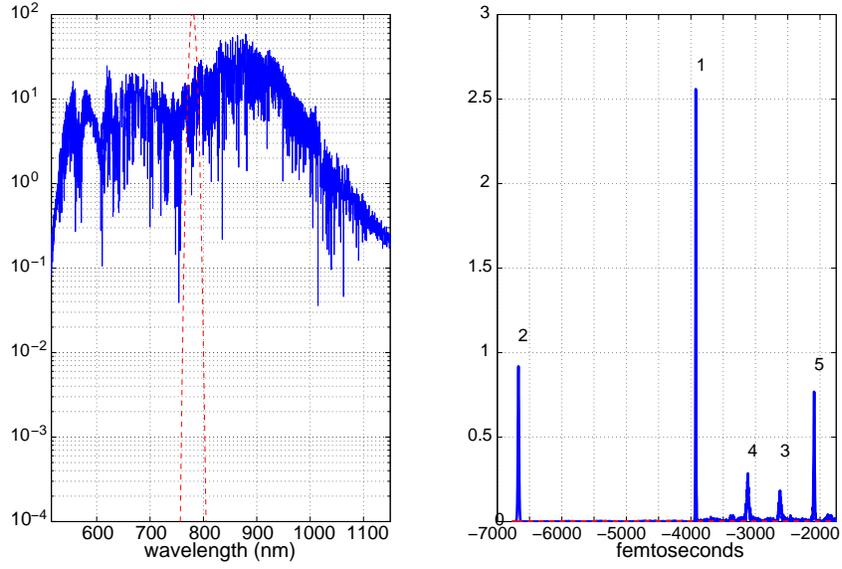}\\
\caption{Spectral (left)
and temporal (right) picture of the output after $1m$ propagation.
All relevant parameters are taken from [5]}
\label{F:figure5}
\end{center}
\end{figure}

\begin{figure}[htp]
\begin{center}
\includegraphics[width=5in]{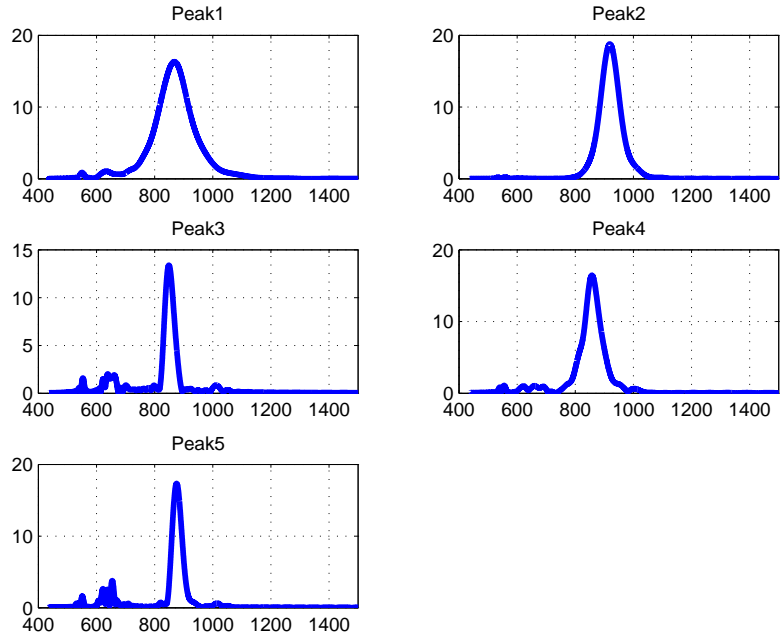}\\
\caption{Spectral characteristics
of the five distinguishable temporal pulses from figure 5.}
\label{F:spectrum1}
\end{center}
\end{figure}

In trying to understand the critical balance between linear and nonlinear effects,
we now depart from the concrete example to further illustrate this interplay in a
series of simulations of the equation shown below, which is no different than equation (7)
except that we placed two adjustable constants $c_2$($c_3$) in front of the self-steepening (linear)
term to account for their respective strengths.

\begin{equation}\label{gNLSE2}
i\partial_z A - c_3 2\pi c{\mathcal F}^{-1}\left(
\int_{\lambda_0}^{\frac{2\pi c}{\omega}} \frac{F(\lambda)}{\lambda^2}
d\lambda \hat{A}(\omega, z)\right) + \gamma\left(1 +
c_2 \frac{i}{\omega_0}\partial_t\right)(A|A|^2)   =0.
\end{equation}
\noindent
By proper re-scaling of the propagation variable $z$ and the pulse peak amplitude, one can
eliminate the parameter $c_3$ but for
clarity we analyze our simulations in the 2-parameter space while  using the same initial condition.
Observe that from (12), an increase of $c_2$
effectively means the shock length is reduced. In practice, this shock length reduction can be induced
by having shorter input pulses.
By allowing ourselves to modulate dispersion and nonlinearity through these
two parameters, we hope to highlight how delicate this balance is. Figure 7 which presents a
separation between two distinct outcomes was obtained by careful
simulations in the $(c_2,c_3)$ parameter space. For the region above the curve the shock is not arrested
and below the curve dispersion regularizes the shock. It is important to point out
the results shown below will strongly depend on the dispersion profile. Nonetheless it is
intriguing to see from figure 7 that a universal critical value $(\frac{c_2}{c_3})_C \approx 2.1$ emerges.
While one could argue that by modifying $c_3$ one departs from a particular photonic structure, what matters
is that for every value $c_3$ (that is moving vertically in figure 7), this transition always occurs.
At this time, we do not have an explanation for it. Furthermore, this property should be tested for different
dispersion profiles.
To summarize, by performing a series of careful numerical simulations where we look at the relative
strengths of the
dispersion (measured by a parameter $c_3$ that multiplies $F(\lambda)$ in
Eq.~(\ref{NLSE})) and of the self-steepening term (measured by a parameter $c_2$) we
clearly demonstrate two dynamical regions: one where the singularity due to the
shock is not suppressed by dispersion (the region above the curve). In
this regime the spectral broadening does not saturate and the numerical solution blows up,
clearly suggesting that additional physical mechanisms must be considered. In the second region
(the region below the curve) propagation leading to
supercontinuum generation or other dynamics similar to those highlighted in the first part of the paper is observed.
Although we did not show the curve beyond $c_3=0.8$, it should be clear that
the point $c_2=c_3=1$ corresponding to the experimental parameters in [3], is as expected below the curve.

\begin{figure}[htp]
\begin{center}
\includegraphics[width=3.5in]{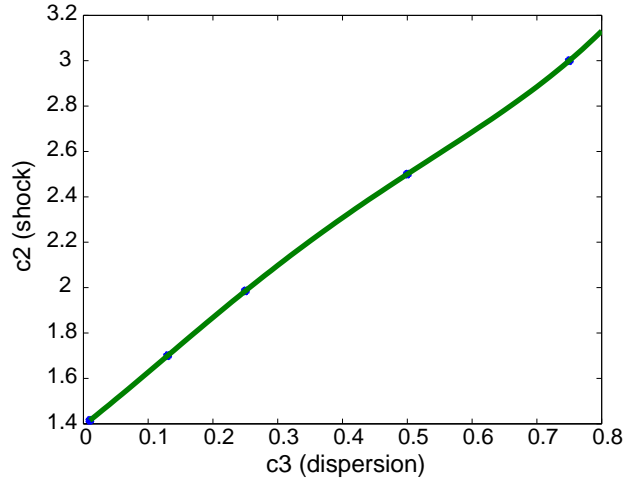}\\
\caption{c2 vs c3 curve that
separates regions where the numerical simulation blows up
(region above the curve)
from the region where supercontinuum is numerically observed (region below the curve).}
\label{F:spectrum}
\end{center}
\end{figure}

\section{SC generation beyond the SVEA}

As research into SC generation extends into regimes exhibiting
the formation of ultra-short temporal pulses, the
validity of the slowly varying envelope approximation (SVEA),
even if we incorporate a detailed dispersion profile over a broad
spectral range, requires careful consideration. Deriving unidirectional propagation
equations for short pulses that would start from Maxwell's equations and account for dynamics such
as higher harmonic generation and sub-cycle shock structure leads to
complex envelope equations which can only be solved numerically \cite{moloney2, optexp07}. One can instead
build computational schemes that solve Maxwell's equations for ultra-short pulses \cite{kinsler}.
Careful integration of generalized envelope equations can relove pulses as shot as sub-50 attoseconds
\cite{optexp07}.
Alternatively, in this section, we illustrate the universality of the features discussed above
and observed in many experiments;
optical shock formation, dispersion regularization and spectral broadening arise, employing models that
do not use the SVEA. Consider the one dimensional nonlinear
wave equation derived from Maxwell's equations,
\begin{eqnarray} \label{maxwell}
\frac{\partial^2A}{\partial z^2}&=&\frac{\partial^2A}{\partial t^2}+
\frac{\partial^2}{\partial t^2}\int \chi^{(1)}(t-\tau)A(z,\tau)d\tau
\label{nonlinwave}\\
\nonumber & +
&\frac{\partial^2}{\partial t^2}\int \chi^{(3)}(t-\tau_1,t-\tau_2,t-\tau_3)A(z,\tau_1)A(z,\tau_2)A(z,\tau_3)d\tau_1 d\tau_2
d\tau_3,
\end{eqnarray}%
where $\chi^{(1)}, \chi^{(3)}$ are linear and nonlinear susceptibilities
(for a detailed derivation of this equation,
see \cite{CS07}).
Due to the characteristics of its derivation, this equation again preserves the nonlocal nature of the
pulse,
and thus is valid for both broad and ultrashort pulses. Moreover, it was reported in \cite{CS07}
that $A(t) = \frac{\alpha}{1+(\beta t)^2}$ is a stationary solution provided
that $\chi^{(3)} = 1$ and the Fourier
transform of linear susceptibility takes the parabolic form,
\begin{equation}\label{chi}
\hat{\chi}^{(1)}(\omega) = -\frac{1}{8}\frac{\alpha^2}{\beta^2}(\omega^2 + 3\beta |\omega| + 3\beta^2),
\end{equation}
for some real constants $\alpha$ and $\beta$. While this may not be a realistic dispersion profile, we
believe it illustrates well  phenomena observed in realistic scenarios.

Here we assume that $\hat{\chi}^{(1)}$ has the form of Eq. (\ref{chi}) with $\alpha =0.2,\beta = 0.75$,
and an initial pulse of $\frac{0.4}{1+0.75^2 t^2}$. Note that the amplitude of the initial pulse is
larger than $\alpha$, and thus the solution will not be stationary and we expect nonlinear
effects will be dominant.

\begin{figure}[ht!]
\begin{center}
\epsfig{figure =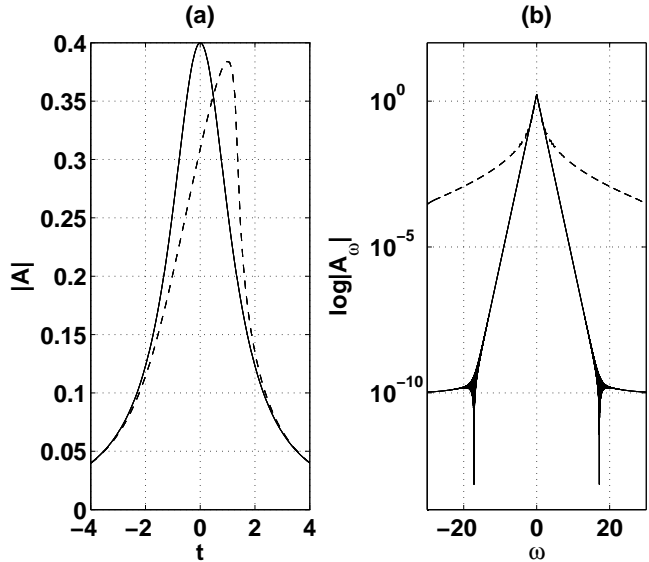, scale=.5}
\caption{(a) The initial pulse (solid) and output pulse (dashed)
after propagation $z= 5$,
resulted from Eq.(\ref{maxwell}), where ${\chi}^{(1)}=0$ and $\chi^{(3)} = 1$.
(b) Spectra of initial pulse (solid) and output pulses (dashed).}\label{nonlinear2}
\end{center}
\end{figure}

\begin{figure}[ht!]
\begin{center}
\epsfig{figure =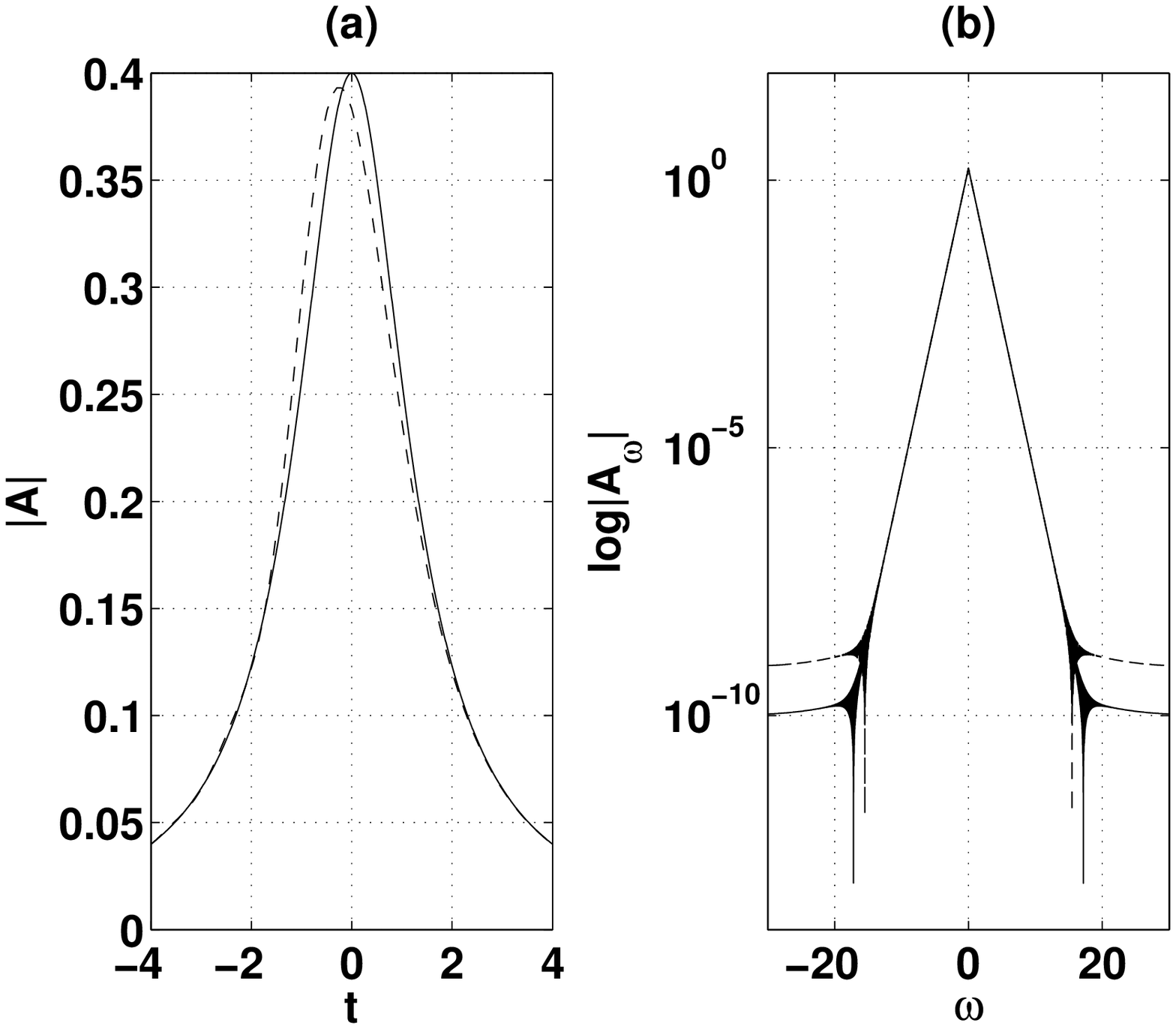,scale=.5}
\caption{(a) The initial pulse (solid) and output pulse (dashed) after propagation $z=5$, resulted
from Eq.(\ref{maxwell}), where $\chi^{(3)} = 0$ and ${\chi}^{(1)}$ as in Eq. (\ref{chi}).
(b) Spectra of initial pulse (solid) and output pulse (dashed).} \label{lin}
\end{center}
\end{figure}

\begin{figure}
\begin{center}
\epsfig{figure=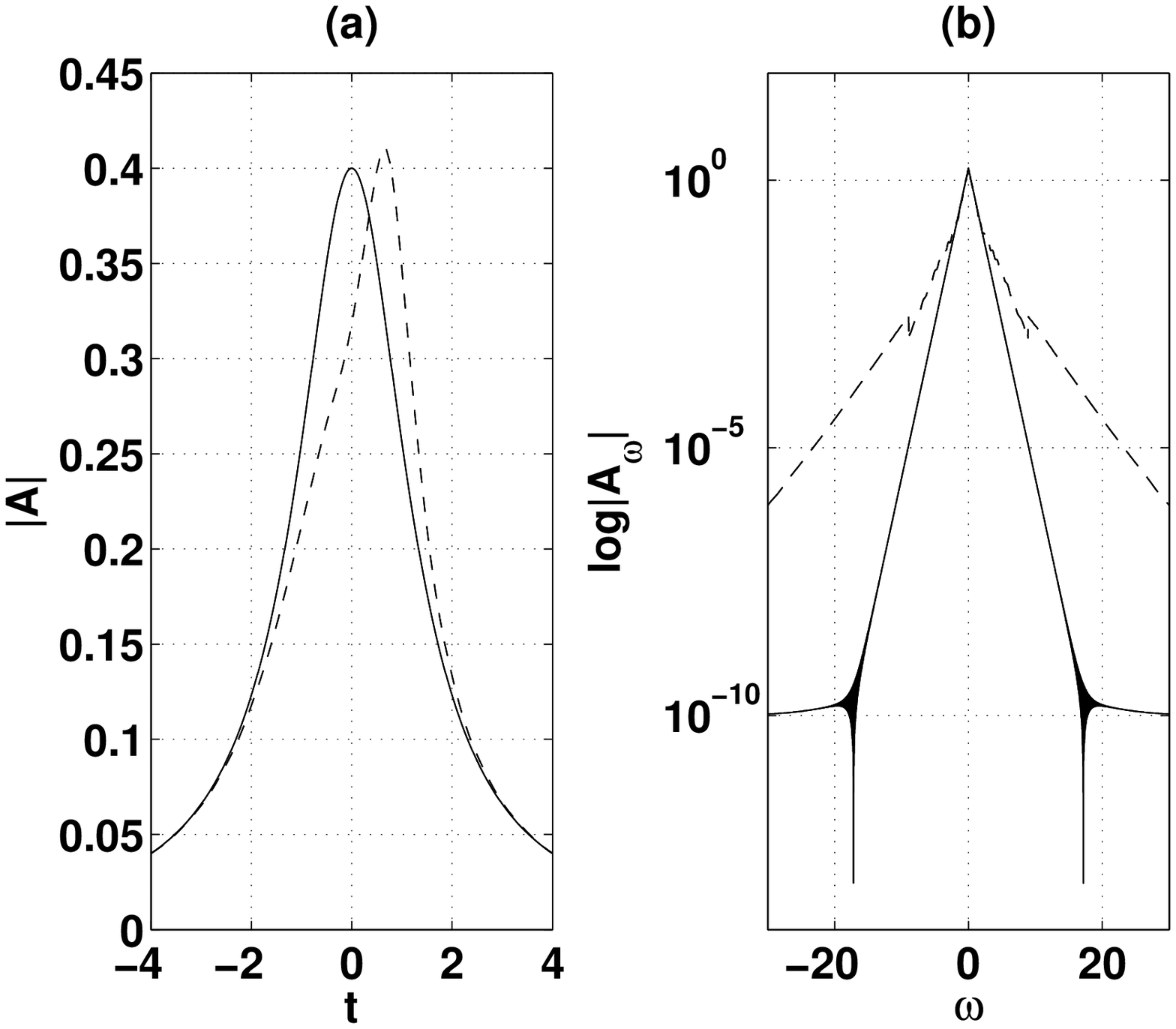,scale=.5}
\caption{(a) The initial pulse (solid) and output pulse (dashed) after propagation $z=5$ with combining
linear and nonlinear effects. The solution is computed based on Eq. (\ref{maxwell}),
where $\chi^{(3)} = 1$ and ${\chi}^{(1)}$ as in Eq. (\ref{chi}). (b) Spectra of initial pulse
(solid) and output pulse (dashed) presented.} \label{combine}
\end{center}
\end{figure}

We first illustrate the role of nonlinear effects by turning off the linear susceptibility, i.e.,
${\chi}^{(1)} = 0,$ and setting the nonlinear coefficient $\chi^{(3)} = 1$. Numerical
simulations of pulse propagation
based on Eq. (\ref{maxwell})
show the output pulse at propagation length $z=5$ and its spectrum (Figure ~\ref{nonlinear2}).
In this case, the nonlinear effects lead to steepening of the pulse shape, which
results in the broad spectrum of the output pulse. In the other extreme,
Fig. ~\ref{lin} illustrates the pure linear effects by turning off the nonlinearity, i.e.,
$\chi^{(3)}=0$, and
using the linear susceptibility as in  (\ref{chi}).
At propagation length $z=5$, the pulse has slightly
changed its form and also its spectrum. Finally, Fig. \ref{combine} presents the combined effects
of both nonlinearity and linearity, i.e., $\chi^{(3)}=1$ and $\chi^{(1)}$ as in (\ref{chi}).
Compared to
Fig.~(\ref{nonlinear2}), the shock is suppressed by linear dispersion. However, with our choice of
input conditions, the nonlinear effect still
dominates the dynamics, which  induces spectral broadening at finite propagation distances.

\section{Conclusions}

Supercontinuum generation is a fascinating and important phenomenon observed in
certain nonlinear wave systems. In this work, we first discussed several simple
models where we tuned dispersion and nonlinearity so that we could showcase different outcomes.
In particular we showed shock driven SC generation as well as soliton-fission driven SC.
Next we moved to a model closely related to an existing
photonic crystal fiber and showed both SC generation as well as critical shock formation.
It is important to emphasize that by properly integrating the dispersive
terms for a given photonic microstructured fiber, we capture
supercontinuum generation as observed in experiments, likely to
greater accuracy than the more common expansion to a finite order of the linear dispersion relation.
Our numerical simulations illustrate that
for some input conditions, shocks rather than soliton fission appear to be dominant
and become the major source of spectral broadening. It is true that soliton fission
as seen in many works
could be the leading mechanism towards SC generation. Which effect is more dominant and what
signatures (if any) of the spectral picture can explain the hierarchy of effects coming into the
dynamics remains unclear. Interestingly, two recent theoretical papers using a wave-turbulence approach \cite{ol08, pra79}
propose an explanation of SC generation. In \cite{ol08}, the authors claim that
coherent structures (solitons, shocks) no
longer play a significant role and instead SC results from a nonequilibrium thermodynamic process.
This was based in an NLSE-type model which included higher order dispersion, but excluded
self-steepening. On the other hand, if self-steepening is included \cite{pra79}, it plays a critical role
in thermalization towards a  two peak SC spectral profile.An experimental signature of such optical thermalization was
recently reported in \cite{optexp09}
Overall, as these works clearly illustrate,
an accurate mathematical model is essential to better reflect
the experimental outcomes. In particular we have a numerical approach at our disposal
to study any photonic fiber structure for which GVD profiles have been or can be computed.

Any time one models SC with a slowly varying envelope approximation the question of validity
comes to mind. In the last section we presented a simple 1D nonlinear Maxwell's equation model
where spectral broadening is achieved, thus re-enforcing the overall principle that SC generation is
universal. While more realistic models that go beyond the SVEA are
more difficult to analyze, careful numerical simulations will lead to a better understanding
of this very rich phenomenon.

\section*{Acknowledgments}
The authors would like to thank the referees for their fair and useful observations and recommendations.

\end{document}